\documentclass[superscriptaddress,prb,twocolumn,showpacs]{revtex4}
\usepackage{amsmath}
\usepackage{amssymb}
\usepackage{graphicx}
\usepackage{bm}
\usepackage{dcolumn}
\usepackage{color}

\begin{document}

\title{Magnetic oscillations measure interlayer coupling in cuprate
superconductors}
\author{P.D. Grigoriev}
\affiliation{L. D. Landau Institute for Theoretical Physics, 142432 Chernogolovka, Russia}
\affiliation{National University of Science and Technology ``MISiS'', Moscow 119049,
Russia}
\affiliation{P.N. Lebedev Physical Institute, RAS, 119991, Moscow, Russia}
\author{Timothy Ziman}
\affiliation{Institut Laue-Langevin, BP 156, 41 Avenue des Martyrs, 38042 Grenoble Cedex
9, France}
\affiliation{LPMMC (UMR 5493), Universit\'e de Grenobles-Alpes and CNRS, Maison des
Magist\`eres, BP 166, 38042 Grenoble Cedex 9, France}

\begin{abstract}
The magnetic oscillations in YBCO high-temperature superconductors have been
widely studied over the last decade and consist of three equidistant low
frequencies with a central frequency several times more intense than its two
shoulders. This remains a puzzle in spite of numerous attempts to explain
the corresponding small Fermi-surface pockets. Furthermore the ARPES data
indicate only four Fermi-arcs with bilayer splitting, and show no sign of
such small areas in the Fermi surface. Here we argue that the magnetic
oscillations measured in under-doped bilayer high temperature
superconductors, in particular YBa$_{2}$Cu$_{3}$O$_{6+\delta }$, provide a
measure of the interplanar electronic coupling rather than the areas of
fine-grain reconstruction of the Fermi surfaces coming from induced charge
density waves. This identification is based on the relative intensities of
the different peaks, as well as their angular dependence, which points to an
effective Fermi surface that is larger than the oscillation frequencies, and
is compatible with several indications from ARPES. The dominance of such
frequencies with respect to the fundamental frequencies from the Fermi
surface is natural for a strongly correlated quasi-two dimensional
electronic systems where non-linear mixings of frequencies are more
resistant to sample inhomogeneity.
\end{abstract}

\date{\today }
\maketitle

\section{Introduction}

Magnetic quantum oscillations (MQO) is a traditional and powerful tool to
study electronic structure of various metals.\cite{Abrik,Shoenberg,Ziman}
The first observation of MQO in cuprate high-temperature superconductors
about a decade ago,\cite{ProustNature2007} was rather a surprise given that
these are normally taken as indicating a Fermi surface of a normal metallic
state. Since then, MQO have been used extensively to investigate the
electronic structure in cuprates, both hole-doped\cite%
{SebastianRepProgPhys2012,ProustComptesRendus2013,AnnuReviewYBCO2015,SebastianPhilTrans2011}
and electron-doped,\cite{HelmNd2009,HelmNd2010,HelmNd2011,HelmNd2015} as
well as in various Fe-based superconductors. \cite%
{BaFeAs2011,Graf2012,ColdeaReview2013,FeSeTerashima2014,FeSeAudouard2015,FeSeWatsonPRB2015,FeSeMQOPRL2015}
Probably the most striking data are for the underdoped yttrium barium copper
oxide (YBCO) compounds YBa$_{2}$Cu$_{3}$O$_{6+\delta }$ (\cite%
{ProustNature2007,SebastianNature2008,AudouardPRL2009,SingletonPRL2010,SebastianPNAS2010,SebastianPRB2010,SebastianPRL2012,SebastianNature2014,ProustNatureComm2015}%
, reviewed in \cite%
{SebastianRepProgPhys2012,ProustComptesRendus2013,AnnuReviewYBCO2015,SebastianPhilTrans2011}%
,where there is one prominent oscillation peak at frequency $F_{\alpha
}\approx 530T$ with two smaller shoulders at $F_{\pm }=F_{\alpha }\pm \Delta
F_{\alpha }$, where $\Delta F_{\alpha }\approx 90T$. All three frequencies
in YBCO are much smaller than expected from closed pockets of any Fermi
surface, seen, for example from ARPES experiments \cite%
{FournierARPES2010,Borisenko2006}.

The completely unreconstructed Fermi surface of YBCO would consist of one
large pocket, almost a square with smoothed corners, filling about one half
of the Brillouin zone and corresponding to a large frequency $\sim 10^{4}$
tesla. Fermi-surface reconstruction, possibly caused by the pseudogap, AFM
or CDW order, takes place for doping level $p<15\%$, resulting to four Fermi
arcs, as suggested by ARPES \cite{FournierARPES2010,Borisenko2006} and
schematically shown in Fig. \ref{FigFS}b. The scattering by the AFM wave
vector $Q=(\pi /a,\pi /b)$, connecting the ends of Fermi arcs, forms closed
FS pockets of area about 6\% of the Brillouin zone, corresponding to MQO
frequency about 1.6kTesla. 
The observed magnetic oscillation frequency $F_{\alpha }\approx 530T$
corresponds to a Fermi-surface cross-section of only $2$\% of the Brillouin
zone, much less than the size of the pockets suggested by ARPES, without
entering into considerations such as whether the "Fermi arcs" can actually
be closed. Thus there is a clear inconsistency between the ARPES and MQO
experimental data. An unusual alternative source of oscillations was
proposed in terms of Andreev-type bound states,\cite{NPhysPereg} but the
predicted change in oscillation frequencies with superconducting gap
contradicts experiment.

The situation has been complicated by the subsequent evidence of at least
fluctuating and short range charge order in low fields by X-ray scattering,%
\cite{XRayScience2012,XRayNatPhys2012,XRayPRL2012,Xray2016} nuclear magnetic
resonance \cite{NMR2011Wu,NMR2015Wu}, and sound velocity measurements \cite%
{CDWSoundVelocity}. In the high magnetic fields corresponding to the range
where magnetic oscillations are observed, superconductivity is gradually
suppressed and charge density wave coherence is stabilized.\cite{Xray2016}
The measured Hall and Seebeck coefficients \cite%
{ElPocketHall2007,BadouxHall2016,ElPocketSeebeck2010}, treated by the
simplified theory without taking into account magnetic breakdown, strong
electronic correlations and superconducting flux flow contribution, support
additional Fermi-surface reconstruction. Thus it is tempting, but we argue
misleading, to attempt to explain the observed low-frequency magnetic
oscillations by the appearance of new Fermi surface pockets coming from
reconstruction of the larger Fermi surfaces by charge density wave order.
Different attempts in this direction vary in details such as inclusion of
spin-orbit or Zeeman splittings,\cite%
{EfimovPRB2008,GarciaNJP2010,HarrisonNJP2012,HarrisonSciRep2015,Briffa2015}
but it is hard to explain the observed three-peak frequency pattern of
quantum oscillations without predicting additional frequencies of similar
amplitudes from the charge order. Note that a frequency pattern somewhat
similar to that of YBa$_{2}$Cu$_{3}$O$_{6+\delta }$ is observed in the
closely related stoichiometric compound YBa$_{2}$Cu$_{4}$O$_{8}$,\cite%
{Yelland2008,Bangura2008,TanPNAS2015} where there is no indication of a
static superstructure. Furthermore if very small Fermi-surface pockets
really are the origin of the observed $F_{\alpha }$, $F_{+}$, and $F_{-}$
frequencies, they should depend strongly on doping,\cite{CommentDoping1}
which does not seem to be the case: for instance $F_{\alpha }$ changes by
only 10\% when the doping $p$ almost doubles, from 0.09 to 0.14.\cite%
{DopingDependence2015}

In this paper, we argue that the magnetic oscillations in these under-doped
bilayer high temperature superconductors in fact provide a measure of the
inter-planar electronic coupling and do \textit{not}, contrary to widespread
belief, correspond to areas of fine-grain reconstruction of the Fermi
surfaces coming from induced charge density waves. This identification is
based on the relative intensities of the different peaks, as well as their
angular dependence, which points to an effective Fermi surface that is
larger than the oscillation frequencies, and is compatible with indications
from ARPES as to the fundamental frequency as well as the bilayer splitting.
The dominance of such frequencies with respect to the standard frequencies
from the Fermi surface, whose current observation is still somewhat
controversial, is natural because the non-linear mixings of frequencies
better survive sample inhomogeneity, as we show in detail below.

\section{Magnetic oscillation produced by interlayer hopping}

\subsection{Qualitative idea}

A clue to the origin of the observed tiny MQO frequencies comes from what
are called \textquotedblleft slow oscillations\textquotedblright\ (SlO) in
organic superconductors \cite{SO}. In these quasi-two dimensional compounds,
oscillations can be clearly attributed, after no little debate, \textit{not}
to new small pockets of the Fermi surface, but to the mixing of two close
frequencies $F_{\beta }\pm \Delta F$, where only the $F_{\beta }$ frequency
corresponds to a Fermi-surface area and the frequency splitting $\Delta F$
is due to Fermi-surface warping, whose origin is the interlayer electron
transfer integral $t_{z}$.\cite{SO} Magnetoresistance (MR) oscillations with
a much lower beat frequency $F_{slow}=2\Delta F=4t_{z}B/\hbar \omega _{c}$
then arise. What is surprising at first sight is that the amplitude of such
emergent slow oscillations is much higher than those of the oscillations
with the original frequencies $F_{\beta }\pm \Delta F$, because they are
damped neither by temperature nor by long-range disorder (spatial
inhomogeneity, leading to variations of $E_{F}$ along the sample on the
scale much larger than magnetic length)\cite{SO,Shub} (see Sec. IID below).
Even in high-quality monocrystals of organic metals such long-range disorder
has been shown to make the major contribution to the Dingle temperature.\cite%
{SO} In the notoriously inhomogeneous cuprates, such disorder is undoubtedly
much stronger, and magnetic oscillations from closed pockets should be even
more strongly damped compared to the beat frequencies.

\begin{figure}[tb]
\includegraphics[width=0.5\textwidth]{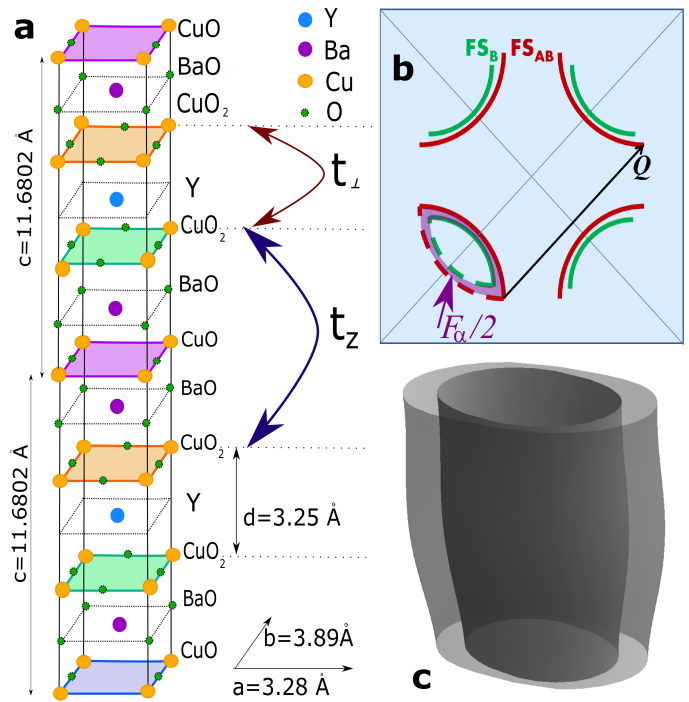}
\caption{\textbf{The bilayer crystal structure and Fermi surface of YBCO. a)}
The crystal structure in YBCO, producing the bilayer splitting to bonding
(B) and antibonding (AB) electron states and Fermi-surface parts. \textbf{b)}
The schematically shown in-plane Fermi surface(FS) in YBCO, seen by ARPES
and without fine-grained reconstruction. The solid green lines show Fermi
arcs of bonding FS, and the red lines show the antibonding FS. The dashed
lines show the FS parts shifted by the vector $Q=(\protect\pi /a,\protect\pi %
/b)$. The dashed and solid lines together form two closed FS pockets,
corresponding to bonding and antibonding states and responsible for the $F_{%
\protect\beta }\sim 1.6kT$ frequency, about $6\%$ of the Brillouin zone. The
most prominent $F_{\protect\alpha }$ frequency $\sim 2\%$ of Brillouin zone
corresponds to double the difference between green and red FS pockets, i.e.
between B and AB FS pockets - the area shaded in purple. \textbf{c)} The
illustration of a quasi-2D Fermi surface with interlayer warping due to $%
2t_{z}$ and double bilayer splitting due to $2t_{\perp }$.}
\label{FigFS}
\end{figure}

To pursue this basic idea, we here extend the theory developed for the
organics to include interplanar couplings from the underlying bilayer
structure of YBCO, illustrated in Fig. \ref{FigFS}. The richer structure
that emerges will, in fact, give a much clearer indication of the origin of
the low-frequency oscillations. In a bilayer structure there are two types
of interlayer hopping: (i) between adjacent layers separated by distance $d$
within one bilayer, given by the transfer integral $t_{\perp }=t_{\perp }(%
\boldsymbol{k}_{\parallel })$, and (ii) between adjacent equivalent bilayers
separated by distance $h$, given by the transfer integral $t_{z}=t_{z}(%
\boldsymbol{k}_{\parallel })$, where $\boldsymbol{k}_{\parallel }$ is the
intralayer momentum. The resulting electron energy spectrum is given by\cite%
{GarciaNJP2010}%
\begin{equation}
\epsilon _{\pm }\left( k_{z},\boldsymbol{k}_{\parallel }\right) =\epsilon
_{\parallel }\left( \boldsymbol{k}_{\parallel }\right) \pm \sqrt{%
t_{z}^{2}+t_{\perp }^{2}+2t_{z}t_{\perp }\cos \left[ k_{z}\left( h+d\right) %
\right] }.  \label{EpsBilayerKz}
\end{equation}%
This electron energy spectrum has, for $t_{z}\ll t_{\perp }$, bonding and
anti-bonding states each with weak $k_{z}$ dispersion and separated by $\sim
2t_{\perp }\left( \boldsymbol{k}_{\parallel }\right) $: 
\begin{equation}
\epsilon _{\pm }\left( k_{z},\boldsymbol{k}_{\parallel }\right) \approx
\epsilon _{\parallel }\left( \boldsymbol{k}_{\parallel }\right) \pm t_{\perp
}\left( \boldsymbol{k}_{\parallel }\right) \pm 2t_{z}\left( \boldsymbol{k}%
_{\parallel }\right) \cos \left[ k_{z}\left( h+d\right) \right] .
\label{EpsBilayerKzApprox}
\end{equation}%
The corresponding in-plane Fermi surface, shown schematically in Fig. \ref%
{FigFS}b, contains four splitted Fermi arcs shown by green lines for bonding
and by red lines for antibonding states, in agreement with ARPES data \cite%
{FournierARPES2010,Borisenko2006}. The dashed lines in Fig. \ref{FigFS}b
denote the Fermi arcs shifted by the vector $Q\approx (\pi /a,\pi /b)$,
which correspond to the reconstructed Fermi surface due to scattering by AFM
or pseudogap ordering. These dashed lines together with solid lines form two
closed Fermi-surface pockets of slightly different area, which may produce
MQO. According to our proposal, the doubled difference between the bonding
(green) and antibonding (red) pockets gives the slow $F_{\alpha }\approx
530T $ MQO frequency observed in YBCO. The two side frequencies come from 3D
warping of these FS pockets, illustrated in Fig. \ref{FigFS}c and
originating from interbilayer electron hopping $t_{z}$, corresponding to the
last term in Eq. (\ref{EpsBilayerKzApprox}). This interpretation is
supported by the observed angular dependence of the $\Delta F_{\alpha }$
splitting \cite{SebastianPRB2010}. Below we give a more quantitative and
detailed substantiation of our interpretation.

\subsection{Analytical formula for slow magnetoresistance oscillations}

According to Eq. (\ref{EpsBilayerKzApprox}), in YBCO there should be at
least two types of splitting of the original frequencies: the larger bilayer
splitting $\Delta F_{\perp }=t_{\perp }B/\hbar \omega _{c}$, where $t_{\perp
}=\left\langle t_{\perp }\left( \boldsymbol{k}_{\parallel }\right)
\right\rangle \neq 0$ and the angular brackets signify an averaging over
in-plane momentum $\boldsymbol{k}_{\parallel }$ on the Fermi surface, and
the smaller splitting $\Delta F_{c}=2t_{z}B/\hbar \omega _{c}\ll \Delta
F_{\perp }\ll F_{\beta }$ due to the $k_{z}$ electron dispersion, where we
also assume $t_{z}=\left\langle t_{z}\left( \boldsymbol{k}_{\parallel
}\right) \right\rangle \neq 0$. These two splittings result in \textit{four}
underlying frequencies $F_{\beta }\pm \Delta F_{\perp }\pm \Delta F_{c}$ of
similar amplitudes. The slow oscillations in magnetoresistance originate
from these \textit{four} frequencies result in a much richer set of
frequencies than for a single layer structures previously considered, which
had only \textit{two} $F_{\beta }\pm \Delta F_{c}$.\cite{SO,RET}

\begin{figure}[tbh]
\includegraphics[width=0.5\textwidth]{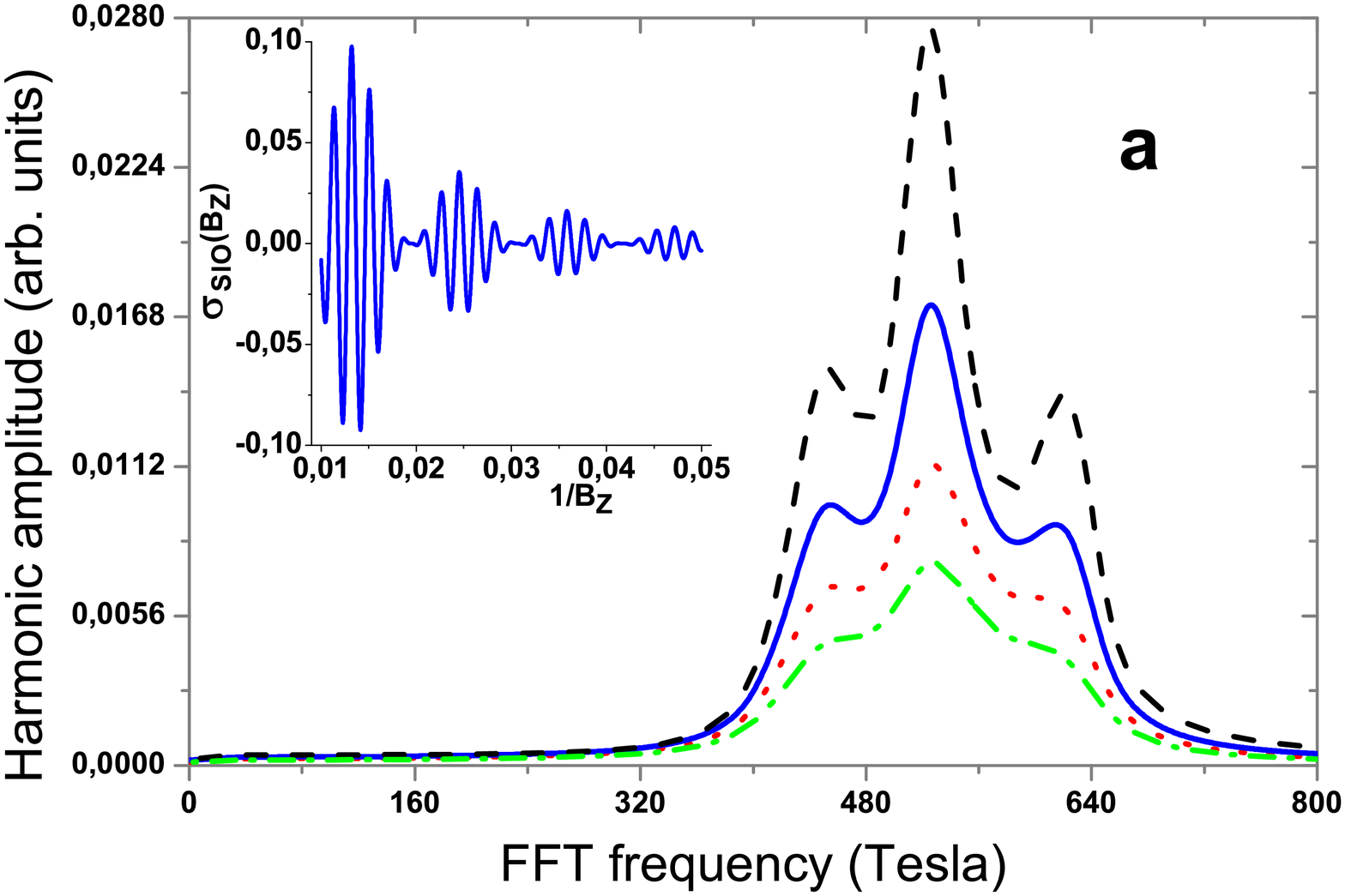} 
\includegraphics[width=0.48\textwidth]{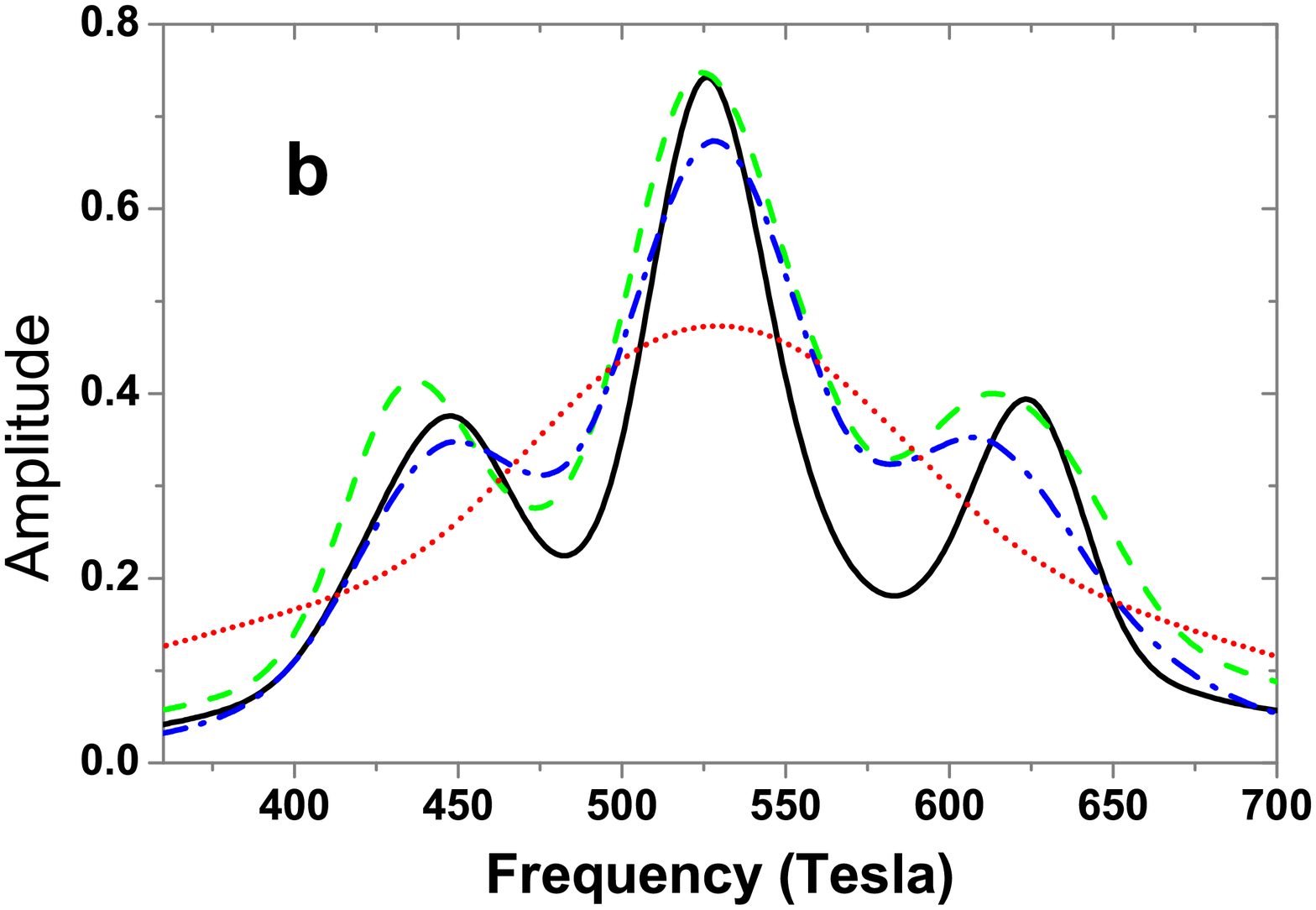}
\caption{\textbf{Predicted magnetic oscillations in quasi-2D conductor with
bilayer splitting.} (a) The Fourier transform of magnetic oscillations given
by Eq. (\protect\ref{sSlO}) at four different Dingle temperatures $\protect%
\pi T_{D}/\hbar \protect\omega _{c}(B_{z}=1T)=1$ (dashed black line), $3$
(solid blue line), $5$ (dotted red line) and $7$ (dash-dotted green line).
Insert shows the initial function $\protect\sigma _{SlO}\left(
1/B_{z}\right) $ at $\protect\pi T_{D}/\hbar \protect\omega _{c}(B_{z}=1T)=3$%
. (b) The Fourier transform of conductivity in Eq. (\protect\ref{sSlO}) at $%
T_{D}=0$ but in the finite field intervals $20T<B_{z}<100T$ (solid black
line), $20T<B_{z}<65T$ (dashed green line), $25T<B_{z}<65T$ (dash-dotted
blue line), and $30T<B_{z}<65T$ (dotted red line).}
\label{Fig2FFT}
\end{figure}

The metallic conductivity along $i$-th axis $\sigma _{i}=\sigma _{ii}$ is
given by the sum of contributions from all ungapped pockets $\beta $ and
over two spin components $s$:%
\begin{equation}
\sigma _{i}=\sum_{\beta ,s}\sigma _{i,\beta ,s}=\sum_{\beta
,s}e^{2}g_{F\beta }D_{i,\beta }.  \label{s}
\end{equation}%
At low temperature each pocket $\beta $ contributes to the total metallic
conductivity along axis $i$ at low temperature via the product of a density
of electron states (DoS) $g_{F,\beta }=g_{\beta }\left( \varepsilon
=E_{F}\right) $ and an electron diffusion coefficient $D_{i,\beta }$. Both
contribute to oscillations, since they vary with the magnetic field $B_{z}$
perpendicular to the conducting $x$-$y$ layers as\cite{Champel2001} (see
Appendix A for a detailed derivation)%
\begin{equation}
\frac{g_{F\beta }}{g_{0\beta }}=1-\sum_{l=\pm 1}2J_{0}\left( 2\pi \frac{%
\Delta F_{c}}{B_{z}}\right) \cos \left( 2\pi \frac{F_{\beta }-l\Delta
F_{\perp }}{B_{z}}\right) R_{D},  \label{g}
\end{equation}%
where $R_{D}=\exp \left( -2\pi ^{2}T_{D}/\hbar \omega _{c}\right) $ is the
Dingle factor\cite{Dingle}, and 
\begin{equation}
\frac{D_{i,\beta }}{D_{0i,\beta }}=1+B_{i,\beta }\sum_{l=\pm 1}J_{0}\left(
2\pi \frac{\Delta F_{c}}{B_{z}}\right) \cos \left( 2\pi \frac{F_{\beta
}-l\Delta F_{\perp }}{B_{z}}\right) R_{D}.  \label{D}
\end{equation}%
Substituting Eqs. (\ref{g}),(\ref{D}) into Eq. (\ref{s}) one obtains four
types of oscillating terms: (i) the first harmonics, of first order in $R_{D}
$ and oscillating rapidly at frequencies $\sim F_{\beta }$; (ii) the second
harmonics with amplitude $\sim R_{D}^{2}$ and frequency $\sim 2F_{\beta }$;
(iii) \textquotedblleft Ultra-slow" oscillations 
\begin{equation}
\sigma _{USlO}\left( B_{z}\right) \propto J_{0}^{2}\left( 2\pi \Delta
F_{c}/B_{z}\right) R_{D}^{2}  \label{uSlO}
\end{equation}
with frequency $\sim 2\Delta F_{c}$; and (iv) \textquotedblleft Slow"
oscillations with frequency $\sim 2\Delta F_{\perp }$: 
\begin{equation}
\sigma _{SlO}\left( B_{z}\right) \propto J_{0}^{2}\left( 2\pi \frac{\Delta
F_{c}}{B_{z}}\right) \cos \left( 4\pi \frac{\Delta F_{\perp }}{B_{z}}\right)
R_{D}^{2}.  \label{sSlO}
\end{equation}%
The Fourier transform (FT) of this magnetic-field dependence of $\sigma
_{SlO}\left( B_{z}\right) $ is shown in Fig. \ref{Fig2FFT}. It closely
resembles the experimental data in YBCO.\cite%
{AnnuReviewYBCO2015,SebastianRepProgPhys2012,SebastianPRB2010}. 
There are three equidistant harmonics, and the amplitudes of the side peaks
relative to the central decrease both with the Dingle factor $R_{D}$ and
with a reduction of the field interval available in experiment.

\subsection{Temperature dependence of slow magnetoresistance oscillations}

At finite temperature $T$ the conductivity tensor is 
\begin{equation}
\sigma \left( T\right) =\int d\varepsilon \,\left[ -n_{F}^{\prime
}(\varepsilon )\right] \,\sigma (\varepsilon ),  \label{sigmayy}
\end{equation}%
where the derivative of the Fermi distribution function $n_{F}^{\prime
}(\varepsilon )=-1/\{4T\cosh ^{2}\left[ (\varepsilon -\mu )/2T\right] \}$,
and $\sigma _{ij}(\varepsilon )$ is the component of the electron
conductivity tensor at energy $\varepsilon $. Usual magnetic oscillation
come from the first-order terms in the Dingle factor, such as $g_{F\beta
}\left( \varepsilon \right) /g_{0\beta }$, which oscillate rapidly as a
function of energy $\varepsilon $: 
\begin{equation}
\frac{\sigma _{1}\left( \varepsilon \right) }{\sigma _{0}}\propto
2J_{0}\left( \frac{4\pi t_{z}}{\hbar \omega _{c}}\right) R_{D}\sum_{\pm
}\cos \left( 2\pi \frac{\varepsilon \pm t_{\perp }}{\hbar \omega _{c}}%
\right) ,  \label{seQO}
\end{equation}%
where $\sigma _{0}$ is the non-oscillating part of conductivity.
Substituting Eq. (\ref{seQO}) into Eq. (\ref{sigmayy}) and performing the
integration over $\varepsilon $, one obtains the well-known result that at
finite temperature the MQO are damped by the factor 
\begin{equation}
R_{T}=\left( 2\pi ^{2}k_{B}T/\hbar \omega _{c}\right) /\sinh \left( 2\pi
^{2}k_{B}T/\hbar \omega _{c}\right) ,  \label{RT}
\end{equation}%
which gives%
\begin{gather}
\frac{\sigma _{1}\left( \mu \right) }{\sigma _{0}}\propto 2J_{0}\left( \frac{%
4\pi t_{z}}{\hbar \omega _{c}}\right) \cos \left( 2\pi \frac{\mu \pm
t_{\perp }}{\hbar \omega _{c}}\right) R_{D}R_{T}  \label{s1} \\
=2J_{0}\left( \frac{2\pi \Delta F_{c}}{B_{z}}\right) \sum_{l=\pm 1}\cos
\left( 2\pi \frac{F_{\beta }-l\Delta F_{\perp }}{B_{z}}\right) R_{D}R_{T}. 
\notag
\end{gather}

The slow oscillations come from the product of oscillating quantities, such
as $\left( g_{F\beta }\left( \varepsilon \right) /g_{0\beta }\right) ^{2}$,
which for bilayer-splitted electron dispersion in Eq. (\ref%
{EpsBilayerKzApprox}) in the second order in $R_{D}$ gives 
\begin{gather}
\sigma _{2}\left( \varepsilon \right) \propto 4J_{0}^{2}\left( \frac{4\pi
t_{z}}{\hbar \omega _{c}}\right) \cos \left( 2\pi \frac{\varepsilon
-t_{\perp }}{\hbar \omega _{c}}\right) \cos \left( 2\pi \frac{\varepsilon
+t_{\perp }}{\hbar \omega _{c}}\right) R_{D}^{2}  \notag \\
=2J_{0}^{2}\left( \frac{4\pi t_{z}}{\hbar \omega _{c}}\right) \left[ \cos
\left( \frac{4\pi \varepsilon }{\hbar \omega _{c}}\right) +\cos \left( \frac{%
4\pi t_{\perp }}{\hbar \omega _{c}}\right) \right] R_{D}^{2}.  \label{se}
\end{gather}%
The first term in the square brackets gives the second harmonic of MQO,
which is strongly damped by temperature by the factor close to the square of 
$R_{T}$ in Eq. (\ref{RT}) similarly to the first harmonic. The second term
in the square brackets of Eq. (\ref{se}), responsible for SlO, is
independent of energy. Therefore, its integration over $\varepsilon $ with $%
n_{F}^{\prime }(\varepsilon )$ in Eq. (\ref{sigmayy}) does not produce any
temperature damping factor for SlO:%
\begin{equation}
\sigma _{2}\left( \mu \right) \propto J_{0}^{2}\left( \frac{4\pi t_{z}}{%
\hbar \omega _{c}}\right) \left[ \cos \left( \frac{4\pi \mu }{\hbar \omega
_{c}}\right) R_{T}^{2}+\cos \left( \frac{4\pi t_{\perp }}{\hbar \omega _{c}}%
\right) \right] R_{D}^{2}.  \label{s2}
\end{equation}%
Thus, within this simplest model, the SlO are not damped by temperature, as
seen in the observed slow oscillations in Refs. \cite{SO,Shub}. However,
magnetoresistance oscillations observed at frequency $F_{\alpha }\approx
530T $ have some temperature damping, corresponding to an effective mass of $%
m^{\star }\approx 1.6m_{e}$.\cite%
{SebastianPRB2010,SingletonPRL2010,CommentFitRT} Such a strong temperature
damping of what are proposed by us as being slow oscillations may arise from
the square of the temperature-dependent Dingle factor $R_{D}$.\cite%
{CommentTD} For non-interacting electrons, when the Dingle factor comes only
from impurity scattering, $R_{D}$ is almost temperature independent. A weak
electron-phonon interaction was also predicted\cite{Engelsberg1970} not to
violate the usual $R_{T}(T)$ to the lowest-order perturbation theory, but
for YBCO at $T\sim 10$K the electron-phonon interaction is not weak, and the
result of Ref. \cite{Engelsberg1970} may not apply. In addition, the
electron-electron interaction, which in cuprates is very strong, gives
considerable temperature dependence to $R_{D}$,\cite{Mirlin} as observed
experimentally and may contribute to the measured effective electron mass $%
m^{\star }\approx 1.6m_{e}$. Note that the extracted effective mass depends
strongly on doping outside the doping interval $0.1<p<0.125$,\cite%
{SingletonPRL2010} and this would reflect the doping dependence of the
strength of e-e interactions.

\subsection{Effect of macroscopic spatial inhomogeneities}

The macroscopic spatial inhomogeneities affect MQO similarly to temperature,
because they smear the Fermi level $\mu $ along the whole sample. To show
explicitely, let us take the most common Gaussian distribution of the
spatially fluctuating shift of Fermi level $\Delta \mu \left( r\right) $,
given by the normalized weight%
\begin{equation}
D\left( \Delta \mu \right) =\left( 1/\sqrt{2\pi }W\right) \exp \left[
-\left( \Delta \mu \right) ^{2}/2W^{2}\right] .  \label{Dmu}
\end{equation}%
In addition to the temperature smearing in Eq. (\ref{sigmayy}), given by the
integration over electron energy $\varepsilon $, conductivity acquires the
coordinate smearing, given by the integration over the shift $\Delta \mu
\left( r\right) $ of chemical potential weighted by Eq. (\ref{Dmu}): 
\begin{equation}
\sigma =\int d\mu \,\sigma (\mu )D\left( \mu -\mu _{0}-\Delta \mu \right) ,
\label{sdmu0}
\end{equation}%
where $\sigma (\mu )$ is given by the sum of first- and second-order terms
given by Eqs. (\ref{s1}) and (\ref{s2}). The first-order terms give the MQO%
\begin{gather}
\frac{\sigma _{1}}{\sigma _{0}}\propto \int d\mu \,D\left( \mu -\mu
_{0}-\Delta \mu \right) 2J_{0}\left( \frac{4\pi t_{z}}{\hbar \omega _{c}}%
\right)  \notag \\
\times \cos \left( 2\pi \frac{\mu \pm t_{\perp }}{\hbar \omega _{c}}\right)
R_{D}R_{T}  \notag \\
=2J_{0}\left( \frac{4\pi t_{z}}{\hbar \omega _{c}}\right) \cos \left( 2\pi 
\frac{\mu _{0}\pm t_{\perp }}{\hbar \omega _{c}}\right) R_{D}R_{T}R_{W}.
\label{s1dmu}
\end{gather}%
Here the last damping factor%
\begin{equation*}
R_{W}=\exp \left( -2\pi ^{2}W^{2}/\hbar ^{2}\omega _{c}^{2}\right)
\end{equation*}%
comes from the spatial variations of the Fermi level. The width $W$ of this
Gaussian distribution of Fermi level contributes to the total Dingle
temperature $T_{D}^{tot}\approx T_{D}^{imp}+T_{D}^{inh}$, where $%
T_{D}^{inh}\approx W$. Note that the Gaussian Fermi-level smearing leads to
the quadratic dependence of $\ln R_{W}$ on magnetic field and on harmonic
number, which was experimentally observed in high-quality samples of
quasi-2D organic metals.\cite{WIPRB2012}

The second-order terms (\ref{s2}) after substitution to Eq. (\ref{sdmu0})
give 
\begin{gather}
\frac{\sigma _{2}}{\sigma _{0}}\propto \int d\mu \,D\left( \mu -\mu
_{0}-\Delta \mu \right) J_{0}^{2}\left( \frac{4\pi t_{z}}{\hbar \omega _{c}}%
\right) R_{D}^{2}  \notag \\
\times \left[ \cos \left( \frac{4\pi \mu }{\hbar \omega _{c}}\right)
R_{T}^{2}+\cos \left( \frac{4\pi t_{\perp }}{\hbar \omega _{c}}\right) %
\right]  \notag \\
=\left[ \cos \left( \frac{4\pi \mu _{0}}{\hbar \omega _{c}}\right)
R_{T}^{2}R_{W}^{4}+\cos \left( \frac{4\pi t_{\perp }}{\hbar \omega _{c}}%
\right) \right] J_{0}^{2}\left( \frac{4\pi t_{z}}{\hbar \omega _{c}}\right)
R_{D}^{2}.  \label{s2dmu}
\end{gather}%
We see, that the second harmonic acquires the enhanced damping factor $%
R_{W}^{4}$ from the Fermi-level variations, while the slow oscillations
remain unchanged.

As a result, the fast quantum oscillations are additionally damped by the
Fermi-level variations, while the SlO are not affected by this type of
disorder.\cite{SO,Shub} Qualitatively, this can be described by a complex
Dingle factor $R_{D}^{tot}=R_{D}R_{W}$, or as a complex Dingle temperature $%
T_{D}^{tot}\approx T_{D}+W$.\cite{SO} The first usual part of the Dingle
factor $R_{D}$ comes from short-range disorder, e.g. impurities, and enters
both the usual fast MQO and SlO. The second part of the Dingle factor $R_{W}$
comes from the smearing of the local Fermi level by macroscopic long-range
inhomogeneities and enters only the fast MQO. In organic metals the second
part of the Dingle temperature $W$ turns out to be unexpectedly large, being
more than four times larger than the first part $T_{D}$, so that $%
T_{D}^{tot}\approx 5.3T_{D}$, as was demonstrated from the experimental data
on the field dependence of the amplitudes of both types of magnetic
oscillations in $\beta $-(BEDT-TTF)$_{2}$IBr$_{2}$.\cite{SO} In cuprates,
which are notoriously inhomogeneous, the difference between $T_{D}^{tot}$
and $T_{D}$ can be even larger. Using Eqs. (\ref{s1dmu}) and (\ref{s2dmu})
and neglecting the second harmonics $\propto \cos \left( 4\pi \mu _{0}/\hbar
\omega _{c}\right) $ in Eq. (\ref{s2dmu}), we can rewrite the conductivity $%
\sigma =\sigma _{1}\left( \mu \right) +\sigma _{2}\left( \mu \right) $ as 
\begin{eqnarray}
\frac{\sigma }{\sigma _{0}} &=&J_{0}\left( \frac{4\pi t_{z}}{\hbar \omega
_{c}}\right) \sum_{l=\pm 1}\cos \left( 2\pi \frac{\mu +2lt_{\perp }}{\hbar
\omega _{c}}\right) R_{T}R_{D}R_{W}  \notag \\
&&+J_{0}^{2}\left( \frac{4\pi t_{z}}{\hbar \omega _{c}}\right) \cos \left( 
\frac{4\pi t_{\perp }}{\hbar \omega _{c}}\right) R_{D}^{2}.  \label{sf}
\end{eqnarray}%
The first line describes usual MQO, probably $F_{\beta }$ frequency in YBCO,
while the second line describes SlO, i.e. $F_{\alpha }$ and $F_{\alpha }\pm
\Delta F$ frequencies in YBCO. The damping factor $R_{T}R_{D}R_{W}$ of usual
MQO is, probably, much sronger than the damping factor $R_{D}^{2}$ of SlO,
which explains why $F_{\beta }$ frequency in YBCO is much weaker and more
fragile than $F_{\alpha }$ and even than $F_{\alpha }\pm \Delta F$
frequencies.

\subsection{Angular dependence of the frequencies of slow magnetoresistance
oscillations}

The angular dependence of the split frequency $2\Delta F_{c}$, proportional
to $t_{z}$, drastically differs from that of $2\Delta F_{\perp }$, related
to the bilayer splitting $t_{\perp }$. The frequency $2\Delta F_{c}\propto
t_{z}$ has a strongly non-monotonic dependence on the tilt angle $\theta $
of the magnetic field:\cite{SO,CommentDFtz} 
\begin{equation}
\Delta F_{c}(\theta )=\Delta F_{c}(\theta =0)J_{0}\left( k_{F}c^{\star }\tan
\theta \right) /\cos \theta ,  \label{Angtz}
\end{equation}%
where $c^{\star }$ ($=11.65\mathring{A}$ for YBCO) is the lattice constant
in the interlayer $z$-direction, and $k_{F}$ is the Fermi momentum. This
angular dependence is the same as for the beat frequency of MQO in a
quasi-2D metal\cite{SO}, but differs strongly from the standard cosine
dependence%
\begin{equation}
F\left( \theta \right) =F\left( \theta =0\right) /\cos \theta ,
\label{AngBylayer}
\end{equation}%
typical for quasi-2D metals where the interplanar coupling is so small that
the Fermi surface can be considered perfectly cylindrical. Eq. (\ref{Angtz})
has an obvious geometrical interpretation\cite{Yam}: the slightly warped
Fermi surface has two extremal cross-sections $S_{ext}$ perpendicular to the
magnetic field $\boldsymbol{B}$, which become equal, to first order in $t_{z}
$, at some tilt angles $\theta _{Yam}$. In contrast $2F_{\perp }$ has the
cosine angular dependence given by Eq. (\ref{AngBylayer}). To see this,
consider the dispersion in Eq. (\ref{EpsBilayerKzApprox}) at $t_{z}\ll
t_{\perp }$, corresponding to the Fermi surface in Fig. \ref{FigFS}c. The
Fermi surface consists of two cylinders along the $z$-axis with base areas $%
S_{even}$ and $S_{odd}$, differing by $\Delta S=S_{even}-S_{odd}=4\pi
t_{\perp }m^{\ast }=const$. In a tilted magnetic field the two corresponding
extremal cross-section areas are $S_{even,odd}=S_{even,odd}/\cos \theta $,
leading directly to Eq. (\ref{AngBylayer}). Thus, according to our
interpretation, the angular dependence of $F_{\alpha }$ frequency from
bilayer splitting obeys the usual cosine dependence given by Eq. (\ref%
{AngBylayer}), while the shift $\Delta F_{\alpha }$ of shoulder frequencies
obeys Eq. (\ref{Angtz}) in agreement with experimental data in Ref. \cite%
{SebastianPRB2010}.

\section{Discussion}

In the previous section we have shown that the interlayer hopping may
produce low-frequency magnetic oscillations, given by Eq. (\ref{sSlO}) and
illustrated in Fig. \ref{Fig2FFT}. Although these slow oscillations contain
usual Dingle factor squared, they may be much stronger than the usual MQO,
because the latter are strongly damped by sample inhomogeneities (see Sec.
IID) or by temperature (see Sec. IIC). The Fourier transform of these slow
oscillations has a natural three-peak structure. The amplitude of the
central frequency $2\Delta F_{\perp }$ is at least twice as large as the
amplitudes of the side frequencies $2\Delta F_{\perp }\pm 2\Delta F_{c}$, as
can be seen from pure combinatorics\cite{JETPL2017MQO} (see Appendix B for
details). As given by Eq. (\ref{sSlO}) and shown in Fig. \ref{Fig2FFT}, the
amplitudes of side peaks are additionally damped by the Dingle factor or due
to the finite field range of available experimental data. This theoretical
plot closely resembles the experimental data in YBCO.\cite%
{AnnuReviewYBCO2015,SebastianRepProgPhys2012,ProustComptesRendus2013,SebastianPRB2010}%
. We therefore propose that the observed \cite%
{ProustNature2007,SebastianNature2008,AudouardPRL2009,SingletonPRL2010,SebastianPRB2010,SebastianPNAS2010,SebastianPRL2012,SebastianNature2014,ProustNatureComm2015}
three equidistant harmonics of magnetic oscillations at low frequency $%
F_{\alpha }\approx 530T$ in YBa$_{2}$Cu$_{3}$O$_{6+\delta }$ (and, probably,
in YBa$_{2}$Cu$_{4}$O$_{8}$ \cite{Yelland2008,Bangura2008,TanPNAS2015}) are
not due to the tiny pockets of the Fermi surface reconstructed by CDW order,%
\cite%
{EfimovPRB2008,GarciaNJP2010,HarrisonNJP2012,HarrisonSciRep2015,Briffa2015,SebastianRepProgPhys2012,ProustComptesRendus2013,AnnuReviewYBCO2015}
but originate from mixing of four frequencies $F_{\beta }\pm \Delta F_{\perp
}\pm \Delta F_{c}$, formed by a fundamental frequency $F_{\beta }$ split by
bilayer and interbilayer electron hopping integrals $t_{\perp }$ and $t_{z}$.
Several terms come from this frequency mixing even in the lowest second
order in Dingle factor. A half of these terms come from various sums of
fundamental frequencies and correspond to the second harmonics of
fundamental frequencies, being much weaker than MQO with fundamental
frequencies appearing in the first order in Dingle factor. The second half
of mixed frequencies correspond to the various differences of fundamental
frequencies, which do not contain electron energy or Fermi level but only
split energies. Hence, these oscillations are robust against sample
inhomogeneities, and therefore are much more pronounced than the MQO with
fundamental frequencies. Such oscillations are responsible for the observed
three-peak Fourier transform. The corresponding analytical expression is
given in Eq. (\ref{sSlO}). 
The frequencies $F_{\alpha }=2\Delta F_{\perp }\approx 530T=2t_{\perp
}B/\hbar \omega _{c}$ and $\Delta F_{\alpha }=2\Delta F_{c}\approx
90T\approx 4t_{z}B/\hbar \omega _{c}$ provide an experimental measure of the
values of bilayer $t_{\perp }$ and inter-bilayer $t_{z}$ average electron
transfer integrals. The extracted bilayer splitting can also be compared to
ARPES data\cite{Borisenko2006}. We summarize arguments for this new
interpretation:

(1) \textit{Frequency pattern.} The multiple extra magnetic oscillation
frequencies predicted from the scenarious with Fermi surface reconstruction
are missing in experiment.\cite%
{EfimovPRB2008,GarciaNJP2010,SebastianPhilTrans2011,HarrisonNJP2012,SebastianRepProgPhys2012,HarrisonSciRep2015,Briffa2015,ProustComptesRendus2013,AnnuReviewYBCO2015}%
. 
In contrast, the proposed model predicts only the three peaks shown in Fig. %
\ref{Fig2FFT}. The only frequency with an amplitude comparable to the side
peaks at $F_{\alpha }\pm \Delta F_{\alpha }$ would be even lower: $2\Delta
F_{c}\sim 90T$. The corresponding analytical expression is given in Eq. (\ref%
{uSlO}). Such ultraslow oscillations could be detected only at low magnetic
field $B_{z}<2\Delta F_{c}$, where they are strongly damped by the Dingle
factor. In addition, these ultraslow oscillations are not exactly periodic
in $1/B_{z}$, especially in high field $B_{z}\sim \Delta F_{c}$, because the
zeros of the Bessel's function in Eq. (\ref{uSlO}) are not exactly
equidistant. Nevertheless, such low-frequency magnetic oscillations 
have been observed recently,\cite{ProustNatureComm2015} 
supporting the proposed interpretation. Our model also predicts a much larger frequency $%
F_{\beta }$ from true Fermi surface pockets but with much smaller amplitude
and much stronger damped by temperature and spatial inhomogeneities.
Experimentally, the $F_{\beta }\approx 1.65$kT frequency was indeed observed
in dHvA \cite{SebastianNature2008} and Tunnel Diode Oscillation cyclotron
resonance\cite{SebastianPNAS2010} measurements, but only at very low
temperatures and not in all samples\cite{SebastianPhilTrans2011}.

(2) \textit{Agreement with ARPES.} The value of the bilayer splitting $%
\Delta _{BS}=0.075\mathring{A}^{-1}$ in momentum space observed by ARPES\cite%
{Borisenko2006} is consistent with the $F_{\alpha }=2\Delta F_{\perp }=530T$
frequency, corresponding to the FS area $S_{\alpha }=0.0507\mathring{A}%
^{-1}\approx 2\%\cdot S_{BZ}$. According to our interpretation it is the
doubled shaded area between the bonding and antibonding Fermi arcs (see Fig. %
\ref{FigFS}b), given by the product $4\Delta _{BS}l_{FA}$, where $%
l_{FA}\approx 0.17\mathring{A}^{-1}$ is the length of a Fermi arc. The value
of the \textit{average} bilayer splitting $t_{\perp }$ expected from
band-structure calculations\cite{Briffa2015,Andersen1995} is also consistent
with $F_{\alpha }$:\cite{Commentm} $t_{\perp }\equiv \left\langle t_{\perp
}\left( \boldsymbol{k}_{\parallel }\right) \right\rangle =\hbar eF_{\alpha
}/2m_{\beta }^{\ast }\approx 8meV$. Note that the maximum value $t_{\perp
}\left( \boldsymbol{k}_{\parallel }\right) $ of bilayer transfer integral
may considerably exceed this average value $t_{\perp }$. Similarly the
observed $t_{z}$-induced splitting $2\Delta F_{c}\approx 90T\approx
4t_{z}B/\hbar \omega _{c}$ gives a reasonable average value $2t_{z}\approx
1.4meV$.

(3) \textit{Doping dependence.} The observed $F_{\alpha }\approx 530T$
depends weakly on the degree of doping \cite%
{SingletonPRL2010,SebastianPNAS2010,DopingDependence2015}, more consistent
with our model than expected\cite{CommentDoping1} from tiny pockets in the
Fermi-surface appearing due to FS reconstruction.

(4) \textit{Damping by sample inhomogeneities.} Long-range spatial
inhomogeneities, common in cuprates, should strongly damp oscillations from
true Fermi-surface pockets due to the smearing of Fermi level similar to the
effect of temperature (see Secs. IIC and IID). This type of disorder affects
the proposed slow oscillations much less, as shown in Sec. IID, making their
observation much easier.

(5) \textit{The angular dependencies of the observed frequencies}\cite%
{SebastianPRB2010} $F_{\alpha }$ and $\Delta F_{\alpha }$ agree with Eqs. (%
\ref{AngBylayer}) and (\ref{Angtz}) correspondingly. The observed strong
angular dependence of the split frequency $\Delta F_{c}(\theta )$ (see Fig.
4a in Ref. \cite{SebastianPRB2010}) is well fit by Eq. (\ref{Angtz}),
suggesting that this frequency is indeed related to the $k_{z}$ electron
dispersion, coming from the interlayer transfer integral $t_{z}$ and giving
the warping of a larger Fermi surface. The extracted\cite{SebastianPRB2010}
first Yamaji angle $\theta _{Yam}\approx 43^{\circ }$ in $\Delta F_{c}\left(
\theta \right) $, corresponding to the first zero of the Bessel function $%
J_{0}\left( k_{F}c^{\star }\tan \theta \right) $, indicates the Fermi
momentum $k_{F}=2.4/c^{\star }\tan \theta _{Yam}=2.2nm^{-1}$ and a true
Fermi-surface area of about $S_{ext}\sim \pi k_{F}^{2}\approx 15nm^{-2}$.
This corresponds to an underlying quantum oscillation frequency of $%
S_{ext}\hbar /2\pi e\approx 1.6$kT, which is far from $F_{\alpha }\approx 530
$T but close to the reported value of $F_{\beta }\approx 1.65$kT \cite%
{SebastianNature2008,SebastianPNAS2010}. 
Note that $F_{\beta }$ was observed only at rather low temperatures and not
in all samples,\cite{SebastianPhilTrans2011} which is consistent with our
model, because frequencies corresponding to real Fermi-surface pockets are
stronger damped by sample-dependent inhomogeneities and by temperature than
slow oscillations.

(6) \textit{Insensitivity to magnetic breakdown.} A CDW-induced Fermi-surface reconstruction is not
necessary for the observation of SlO. Hence, there are no issues to be
resolved as how a weak and fluctuating CDW ordering, with transition
temperature less than $40$K, overcomes the magnetic breakdown, which should
be strong for fields up to 100 tesla and restore the electron orbits of
unreconstructed Fermi surface.

(7) \textit{The relative amplitudes of the observed frequencies} are
naturally explained without additional fitting parameters (see Fig. \ref%
{Fig2FFT}).

Very similar magnetic oscillations have been observed in another closely
related bilayer stoichiometric high-Tc superconductor YBa$_{2}$Cu$_{4}$O$%
_{8} $.\cite{Yelland2008,Bangura2008,TanPNAS2015} This compound has no
indication of a CDW order, which excludes the CDW-induced fine-grained FS
reconstruction giving small FS pockets. Hence, the proposed magnetic
oscillations due to bilayer splitting is a quite general phenomenon. Not
only cuprates but also rare-earth tritelluride compounds with bilayer
crystal structure show similar slow oscillations\cite{RET} originating from
bilayer splitting, but the three-peak structure is not resolved there
because of a very small value of $t_z$. Similar slow oscillations with a
frequency 
$\approx 840T$ (but without side-frequency splitting), corresponding to 3\%
of the Brillouin zone, have been observed in HgBa$_{2}$CuO$_{4+\delta }$,
where there is no bilayer splitting, but a much larger $t_{z}$ is expected 
from the shorter interlayer distance. It is as yet unclear whether this
frequency has the same origin, coming from $t_z$, spin-orbit, magnetic or
some other type of splitting. Therefore, a more detailed study of these
magnetic oscillations, as well as the further theoretical study of this
effect taking into account various types of electron-band splitting is need.

To summarize, we propose and analyze a new interpretation of the observed
magnetic oscillations in YBa$_{2}$Cu$_{3}$O$_{6+\delta }$ high-Tc
superconductors, with frequencies $F_{\alpha }\approx 530T$ and $F_{\alpha
}\pm \Delta F_{\alpha }$ related to the bilayer splitting and $t_z$%
-dispersion rather than to tiny Fermi-surface pockets. This is based on the
new result (\ref{sSlO}), as illustrated in Fig. \ref{Fig2FFT}. The
frequencies agree with ARPES data, thus resolving the long-standing puzzle,
and allow us to estimate values of bilayer splitting $t_{\perp }=\hbar
eF_{\alpha }/2m^{\ast }$ and of $k_{z}$-dispersion $t_{z}=\hbar e\Delta
F_{\alpha }/4m^{\ast }$. The observed angular dependence of $\Delta
F_{\alpha }$ points to a true Fermi-surface pocket close to $F_{\beta
}\approx 1.6kT$. This interpretation is robust, requiring only a bilayer
crystal structure and closed Fermi-surface pockets. The reproducibility of $%
F_{\alpha }\approx 530T$ and $F_{\alpha }\pm \Delta F_{\alpha }$ frequencies
in YBCO, in contrast to that of the more fragile $F_{\beta }\approx 1.6kT$,
is understood in terms of the lesser sensitivity to sample inhomogeneities.
The comparison of the amplitudes of $F_{\alpha }$ and $F_{\beta }$
oscillations then provides a potentially useful way of characterizing the
type of disorder present in a given sample. Verifying the existence, and, in
principle a further splitting of the $F_{\beta }$, in extremely clean
samples would be a good test of the proposed theory.

This brings us back to the issue of the charge density order, where we have
mentioned there is now mounting evidence, particularly at the high fields
where magnetic oscillations are observed.\cite{Xray2016} While we have
argued that the magnetic oscillations that were seen \textit{should not} be
taken as evidence of any corresponding detailed Fermi surface
reconstruction, we cannot rule out the possibility of such reconstruction.
Any new oscillations would be subject to magnetic breakdown, which depends
on the strength of charge density order and the disorder. It would be
important to quantify both aspects fully, by detailed study of charge order
and magnetic oscillation as a function of doping and applied pressure, but
it is likely that any new oscillations would be more fragile than the
oscillations we describe. In our view, the bilayer geometry of YBCO is ideal
in identifying the source of the oscillations that have been seen, but it
could well be that in single-layer cuprates a similar mechanism can explain
low frequency components.

The proposed model not only resolves the puzzle of inconsistency between
ARPES and MQO data, but also suggests how the available extensive
experimental data on MQO in YBCO can be used. According to our
interpretation, the frequency of the observed slow magnetic oscillations in
YBCO gives the area between bonding and antibonding Fermi-surface arcs much
more precisely than can be obtained from ARPES data.\cite{Borisenko2006} The
angular dependence of the observed shoulder frequencies allows to estimate
the areas of the true Fermi-surface pockets, which may stimulate their
reliable experimental observation. The comparison of the amplitudes of SlO
and true MQO would give the important information about the spatial
variations of the Fermi level due to sample inhomogeneities. Our model also
predict the magnetic oscillations with ultra-low frequency $\sim 90$ tesla,
which has an unusual angular dependence given by Eq. (\ref{Angtz}). Possibly
such oscillations were detected in Ref. \cite{ProustNatureComm2015}, but the
angular dependence of their frequency needs further experimental study.

\acknowledgments{The authors thank Ted Forgan for raising this
issue and communication of unpublished results. The work
is partially supported by the Ministry of Education and Science
of the Russian Federation in the framework of Increase
Competitiveness Program of MISiS and by RFBR. 
P.G. acknowledges the RSCF grant \# 16-42-01100.}

\appendix

\section{Density of states in quasi-2D metal with bilayer crystal structure}

In a single-band quasi-2D metal without bilayer splitting with only one
cylindrical Fermi Surface, an electron dispersion in the tight-binding
approximation is\cite{Abrik} 
\begin{equation*}
\epsilon \left( k_{z},\boldsymbol{k}_{\parallel }\right) \approx \epsilon
_{\parallel }\left( \boldsymbol{k}_{\parallel }\right) -2t_{z}\cos \left(
k_{z}c^{\star }\right) .
\end{equation*}%
In magnetic field $\boldsymbol{B}$ the electron spectrum consists of a set
of equidistant Landau levels separated by $\hbar \omega _{c}=\hbar
eB_{z}/m^{\ast }c$, where $m^{\ast }$ is the effective electron mass: 
\begin{equation*}
\epsilon _{s}\left( k_{z},n\right) \approx \hbar \omega _{c}\left(
n+1/2\right) -2t_{z}\cos \left( k_{z}c^{\star }\right) +s\Delta _{Z},
\end{equation*}%
where $\Delta _{Z}=\mathrm{g}\mu _{B}B$ is the Zeeman splitting, $\mathrm{g}$
is the g-factor, $\mu _{B}$ is the Borh magneton, and $s=\pm 1/2$ is the
projection of quasiparticle spin on magnetic field $\boldsymbol{B}$,
inclined by the angle $\theta $ with respect to the normal to conducting
layers. The corresponding oscillating density of electron states (DoS) $%
g\left( \varepsilon \right) $ is given by the sum over Landau levels (LL): 
\begin{gather}
g\left( \varepsilon \right) =\sum_{n=0}^{\infty }\sum_{k_{z},s}\delta \left[
\varepsilon -\epsilon _{s}\left( k_{z},n\right) \right]  \notag \\
=\sum_{n=0}^{\infty }\sum_{s=\pm 1}\frac{g_{LL}/\pi }{\sqrt{%
4t_{z}^{2}-\left( \varepsilon -s\Delta _{Z}-\hbar \omega
_{c}\,(n+1/2)+i\Gamma \right) ^{2}}},  \label{gq2D0}
\end{gather}%
where the LL degeneracy per unit area $g_{LL}=eB_{z}/2\pi \hbar c$, and $%
\Gamma =\Gamma \left( \varepsilon \right) $ is the broadening of electron
levels due to scattering. It is often convenient to transform this sum over
LL to the sum over harmonics using the Poisson summation formula. When MQO
are weak, e.g. at $\Gamma \gtrsim \hbar \omega _{c}$, keeping only the
zeroth and first harmonics one obtains:\cite{Champel2001}%
\begin{equation}
\frac{g\left( \varepsilon \right) }{g_{0}}=1-2J_{0}\left( \frac{4\pi t_{z}}{%
\hbar \omega _{c}}\right) \cos \left( \frac{2\pi \varepsilon }{\hbar \omega
_{c}}\right) \cos \left( \frac{2\pi \Delta _{Z}}{\hbar \omega _{c}}\right)
R_{D},  \label{rho0}
\end{equation}%
where $g_{0}=m^{\ast }/\pi \hbar ^{2}d=2g_{LL}/\hbar \omega _{c}d$ is the
average DoS at the Fermi level per two spin components, $J_{0}\left(
x\right) $ is the Bessel's function, and $R_{D}=\exp \left( -2\pi \Gamma
/\hbar \omega _{c}\right) $ is the Dingle factor.\cite{Shoenberg} The DoS at
the Fermi level is found by making the subsititution $\varepsilon
\rightarrow \mu $, where $\mu $ is the chemical potential equal to Fermi
energy. Introducing the magnetic quantum oscillation frequency $F_{\beta
}=\mu B_{z}/\hbar \omega _{c}$, we rewrite Eq. (\ref{rho0}) as 
\begin{equation}
\frac{g_{F}}{g_{0}}=1-2J_{0}\left( \frac{2\pi \Delta F_{c}}{B_{z}}\right)
\cos \left( 2\pi \frac{F_{\beta }}{B_{z}}\right) \cos \left( \frac{2\pi
m^{\ast }}{m\cos \theta }\right) R_{D}.  \label{rho1}
\end{equation}%
For conductivity the sum over two spin components appears only in the final
expression in Eq. (\ref{s}), which contains the DoS per one spin component
without factor $\cos \left( 2\pi m^{\ast }/m\cos \theta \right) $. Below we
do not keep the Zeeman spin splitting $\Delta _{Z}$ and the spin subscript $%
s $ in the intermediate formulas, but similar to the standard calculations%
\cite{Abrik,Shoenberg,Ziman} introduce it only in the end as an energy shift
by $\pm \Delta _{Z}$.

In bilayer crystals the quasi-2D electron dispersion (\ref%
{EpsBilayerKzApprox}) is additionally split by $2t_{\perp }\left( 
\boldsymbol{k}_{\parallel }\right) $ to the so-called "bonding" and
"atibonding" states. The corresponding DoS for one spin component is given
by the sum of contributions from the bonding and antibonding branches of
electron dispersion%
\begin{equation}
g\left( \varepsilon \right) =\sum_{n=0}^{\infty }\sum_{\pm }\frac{g_{LL}/\pi 
}{\sqrt{4t_{z}^{2}-\left( \varepsilon \pm t_{\perp }-\hbar \omega
_{c}\,(n+1/2)+i\Gamma \right) ^{2}}}.
\end{equation}%
Applying the Poisson summation formula and keeping only the zeroth and first
harmonics, similar to Eq. (\ref{rho0}), one obtains:%
\begin{equation}
\frac{g\left( \varepsilon \right) }{2g_{0}}=1-\sum_{l=\pm 1}2J_{0}\left( 
\frac{4\pi t_{z}}{\hbar \omega _{c}}\right) \cos \left( \frac{2\pi
\varepsilon -lt_{\perp }}{\hbar \omega _{c}}\right) R_{D},
\end{equation}%
which gives the DoS on the Fermi level $\varepsilon =\mu $, corresponding to
Eq. (\ref{g}). 
\begin{equation}
\frac{g_{F}}{2g_{0}}=1-\sum_{l=\pm 1}2J_{0}\left( \frac{2\pi \Delta F_{c}}{%
B_{z}}\right) \cos \left( 2\pi \frac{F_{\beta }-l\Delta F_{\perp }}{B_{z}}%
\right) R_{D}.  \label{gq2db}
\end{equation}

The diffusion coefficient $D_{i}$ depends on its direction $i$ with respect
to magnetic field and to the crystal axes, and its oscillations can be
approximated by 
\begin{equation}
\frac{D_{i}}{D_{0i}}\approx 1-B_{i}J_{0}\left( \frac{2\pi \Delta F_{c}}{B_{z}%
}\right) \sum_{l=\pm 1}\cos \left( 2\pi \frac{F_{\beta }-l\Delta F_{\perp }}{%
B_{z}}\right) R_{D},  \label{D0}
\end{equation}%
where $D_{0i}$ is the average non-oscillating part of diffusion coefficient
and the coefficients $B_{i}\sim 1$ depend on the direction $i$. For $%
i\parallel z$ and without bilayer splitting it was calculated in Refs. \cite%
{SO,Shub}, where in addition to $J_{0}\left( 4\pi t_{z}/\hbar \omega
_{c}\right) $ the terms containing $J_{1}\left( 4\pi t_{z}/\hbar \omega
_{c}\right) $ appear, which only introduce the phase shift of beats and of
slow oscillations. For $i\perp z$ these $J_{1}\left( 4\pi t_{z}/\hbar \omega
_{c}\right) $\ terms are small by a factor $\sim t_{z}/E_{F}$,\cite%
{GrigMogInPlaneMR} and Eq. (\ref{D0}) is valid.

\section{Combinatorical explanation of the observed 3-peak structure of MQO}

The observed 3-peak structure of MQO Fourier transform, with the central
peak at least twice larger than the side peaks, can be obtained from simple
combinatorics without using the Bessel's functions. The combination of
double bilayer splitting by $t_{\perp }\left( \boldsymbol{k}_{\parallel
}\right) $ and the Fermi-surface warping due to $t_{z}$ with two extremal FS
cross-section areas gives a four-fold splitting of MQO frequency $F$
corresponding to the true FS pocket area both for the DoS $g_{F}$ and for
the diffusion coefficient $D_{i}$ oscillate with varing of magnetic field $%
B_{z}$ perpendicular to the conducting $x$-$y$ layers as: 
\begin{equation}
\frac{g_{F}}{g_{0}}=1+A\sum\limits_{j,l=\pm 1}\cos \left( 2\pi \frac{%
F_{\beta }+j\Delta F_{\perp }+l\Delta F_{c}}{B_{z}}\right) ,  \label{g0F}
\end{equation}%
and 
\begin{equation}
\frac{D_{i}}{D_{0i}}=1+B_{i}\sum\limits_{j,l=\pm 1}\cos \left( 2\pi \frac{%
F_{\beta }+j\Delta F_{\perp }+l\Delta F_{c}}{B_{z}}\right) ,  \label{D0F}
\end{equation}%
The product of Eqs. (\ref{g0F}) and (\ref{D0F}) gives 
\begin{eqnarray}
\frac{\sigma _{i}}{\sigma _{i}^{(0)}} &=&1+\left( A+B_{i}\right)
\sum\limits_{j,l=\pm 1}\cos \left( 2\pi \frac{F_{\beta }+j\Delta F_{\perp
}+l\Delta F_{c}}{B_{z}}\right)  \notag \\
&&+A_{\beta }B_{i}\sum\limits_{j,l,j^{\prime },l^{\prime }=\pm 1}\cos \left(
2\pi \frac{F_{\beta }+j\Delta F_{\perp }+l\Delta F_{c}}{B_{z}}\right) \times
\notag  \label{Slow0} \\
&&\times \cos \left( 2\pi \frac{F_{\beta }+j\,j^{\prime }\Delta F_{\perp
}+l\,l^{\prime }\Delta F_{c}}{B_{z}}\right) ,  \label{Slow1}
\end{eqnarray}%
where $\sigma _{i}^{(0)}=e^{2}g_{0}D_{0i}$ does not oscillate. The second
term in Eq. (\ref{Slow1}) gives the first harmonic of MQO with amplitudes $%
\sim A$ and four close frequencies $F_{\beta }\pm \Delta F_{\perp }\pm
\Delta F_{c}\sim F_{\beta }\gg \Delta F_{\perp }$, corresponding to one
splitted FS pocket. The last term in Eq. (\ref{Slow1}) is of the second
order in the amplitude $A_{\beta }$ and gives various frequencies: (i) for $%
j^{\prime }=l^{\prime }=1$ the 4 second harmonics $2\left( F_{\beta }\pm
\Delta F_{\perp }\pm \Delta F_{c}\right) $, which are strongly damped by
temperature and disorder and can be neglected; (ii) for $j^{\prime }=1$ and $%
l^{\prime }=-1$ the SlO with ultra-low frequency $2\Delta F_{c}\ll \Delta
F_{\perp }\ll F_{\beta }$, corresponding to the very low frequency $\sim
100T $ recently observed\cite{ProustNatureComm2015} in YBCO; (iii) for $%
j^{\prime }=-1$ and $l^{\prime }=\pm 1$ the SlO with intermediate
frequencies $2\Delta F_{\perp }$ and $2\Delta F_{\perp }\pm 2\Delta F_{c}$,
which correspond to the observed $F_{\alpha }\approx 530T$ and $F_{\pm
}=F_{\alpha }\pm \Delta F_{\alpha }$ frequencies in YBCO. Indeed, using the
identity $2\cos x\cos y=\cos \left( x-y\right) +\cos\left( x+y\right) $ and
neglecting the high frequency ($\sim 2F_{\beta }$) contributions $\cos
\left( x+y\right) $ we can rewrite the last term in Eq. (\ref{Slow1}) for $%
j^{\prime }=-1$ as 
\begin{gather}
\frac{A_{\beta }B_{i,\beta }}{2}\sum\limits_{j,l,l^{\prime }=\pm 1}\cos
\left( 2\pi \frac{2j\Delta F_{\perp }+l\left( 1-l^{\prime }\right) \Delta
F_{c}}{B_{z}}\right) =  \notag \\
A_{\beta }B_{i,\beta }\left[ 2\cos \left( \frac{4\pi \Delta F_{\perp }}{B_{z}%
}\right) +\sum\limits_{l=\pm 1}\cos \left( 4\pi \frac{\Delta F_{\perp
}+l\Delta F_{c}}{B_{z}}\right) \right] .  \label{3h}
\end{gather}%
The first term in the square brackets, corresponding to $l^{\prime }=1$,
gives the central peak, while the last term, corresponding to $l^{\prime
}=-1 $, gives two side peaks. The amplitude of the central peak is doubled
because there are four different combinations of $j,l$ giving this term: any 
$j$ and $l$ at $l^{\prime }=1$. On contrary, each side peak is given by only
two combinations of $j,l$ at $l^{\prime }=-1$: $j=l$ for the right side
peak, and $j=-l$ for the left side peak. This combinatoric analysis was
proposed in Ref. \cite{JETPL2017MQO}.

\section{Possible appearance of slow oscillations in magnetization}

Similarly one can explain the fact that the slow oscillations with frequency 
$F_{\alpha }\approx 530T$ are observed also in dHvA effect, which again
contradicts the simple model above and that of Refs. \cite{SO,Shub}. If the
electron-electron interaction is included, its effects are roughly
proportional to the product of the oscillating density of states (DoS), and
it should produce a nonlinearity in the magnetic oscillations of
magnetization (or of other thermodynamic quantities) as a functional of the
density of states. This nonlinearity results in slow oscillations of
thermodynamic quantities such as magnetization, similar to those of the
magneto-resistance.

Slow oscillations originate from mixing of different but close fundamental
frequencies of magnetic oscillations. This mixing requires some
nonlinearity. Transport properties naturally contain such nonlinearity,
because they are determined by the density of states and diffusion
coefficient. However, the thermodynamic potential%
\begin{equation}
\Omega (\mu ,B)=-k_{B}T\int d\varepsilon \rho (\varepsilon ,B)\ln \left(
1+\exp \frac{\mu -\varepsilon }{k_{B}T}\right) \ ,  \label{Om}
\end{equation}%
is a linear functional of the density of states (DoS) $\rho (\varepsilon ,B)$%
. Hence, thermodynamic quantities, e.g. magnetization $M=-\partial \,\Omega
(\mu ,B)/\partial B$, are also linear functionals of the density of states
(DoS) $\rho (\varepsilon ,B)$.\cite{CommentChemPotOsc} Hence, for slow
oscillations to take place in magnetization, the non-linearity must appear
already in the DoS $\rho (\varepsilon ,B)$. The DoS is related to the
imaginary part of the retarded Green's function $G_{R}(\eta ,\varepsilon )$
as $\rho (\varepsilon ,B)=-\sum_{\eta }\mathrm{Im}G_{R}(\eta ,\varepsilon
)/\pi $, where $\eta =\left\{ n,k_{y},k_{z}\right\} $ is a set of quantum
numbers of quasiparticles in magnetic field. This Green's function $%
G_{R}(\eta ,\varepsilon )$ includes all relevant types of interaction: with
impurities, electron-phonon and electron-electron, which are contained in
the self-energy part $\Sigma _{R}\equiv \Sigma _{R}(\eta ,\varepsilon ,B)$: $%
G_{R}(\eta ,\varepsilon )=\left[ \varepsilon -\epsilon (\eta ,B)-\Sigma _{R}%
\right] ^{-1}$, where $\epsilon (\eta ,B)$ is the bare energy spectrum in
the presence of magnetic field. Hence, the DoS 
\begin{widetext}
\begin{equation}
\rho (\varepsilon ,B)=\sum_{\eta }\frac{-{\rm Im}\Sigma _{R}(\eta
,\varepsilon ,B)/\pi }{\left[ \varepsilon -\epsilon (\eta ,B)-{\rm Re}%
\Sigma _{R}(\eta ,\varepsilon ,B)\right] ^{2}+\left[ {\rm Im}\Sigma
_{R}(\eta ,\varepsilon ,B)\right] ^{2}}.  \label{rhoSigma}
\end{equation}%
If the self-energy part $\Sigma (\eta ,\varepsilon ,B)$ is independent of $%
\eta $, the sum over $\eta $ in Eq. (\ref{rhoSigma}) can be calculated
analytically. In quasi-2D metals $\epsilon
(\eta ,B)=\hbar \omega _{c}\left( n+\gamma \right) +2t_{z}\cos \left(
k_{z}d\right) $, and applying the Poisson summation formula one obtains\cite{Champel2001,Shub}
\begin{equation}
\frac{\rho (\varepsilon ,B)}{\rho _{0}} = 1+2\sum_{k=1}^{\infty
}(-1)^{k}J_{0}\left( \frac{4\pi kt_{z}}{\hbar \omega _{c}}\right)
\exp \left( -2\pi k\frac{\left\vert {\rm Im}\Sigma
_{R}(\varepsilon ,B)\right\vert }{\hbar \omega _{c}}\right)  \cos
\left( 2\pi k\frac{\varepsilon -{\rm Re}\Sigma _{R}(\varepsilon
,B)}{\hbar \omega _{c}}\right) \, .
\end{equation}%
\end{widetext}

The non-linearity of the oscillating field dependence is clearly seen from
this expression. Naively, one would expect that already the impurity
scattering, which in the Born approximation gives oscillating $\mathrm{Im}%
\Sigma _{R}(\varepsilon ,B)\propto \rho (\varepsilon ,B)$, results to slow
oscillations of $\rho (\varepsilon ,B)$. However, this is not the case,
because, as first shown in Ref. \cite{Shub}, in the lowest (second) order of
the Dingle factor $\exp \left( -2\pi \left\vert \mathrm{Im}\Sigma
_{R}(\varepsilon ,B)\right\vert /\hbar \omega _{c}\right) $, the
contribution from $\mathrm{Re}\Sigma _{R}$ exactly cancels the contribution
from $\mathrm{Im}\Sigma _{R}(\varepsilon ,B)$ to the slow oscillations. Of
course, if one goes beyond the Born approximation and/or includes also the
e-e interaction, this exact cancellation does not take place, which results
to slow oscillations of magnetization. Therefore, in organic metals, which
are rather good clean metals, the slow oscillations of magnetization were
not observed.\cite{SO} However, in cuprates, where e-e interaction is very
strong and the impurity concentration is very high, the slow oscillations of
magnetization should appear.


\begin{thebibliography}{99}
\bibitem{Abrik} A.A. Abrikosov, \textit{Fundamentals of the theory of metals}%
, North-Holland, 1988.

\bibitem{Shoenberg} Shoenberg D. \textit{\ Magnetic oscillations in metals},
Cambridge University Press 1984.

\bibitem{Ziman} J. M. Ziman, \textit{Principles of the Theory of Solids},
Cambridge Univ. Press 1972.

\bibitem{ProustNature2007} Nicolas Doiron-Leyraud, Cyril Proust, David
LeBoeuf, Julien Levallois, Jean-Baptiste Bonnemaison, Ruixing Liang, D. A.
Bonn, W. N. Hardy, Louis Taillefer, Nature \textbf{447}, 565 (2007).

\bibitem{AnnuReviewYBCO2015} Suchitra E. Sebastian and Cyril Proust, Annu.
Rev. Condens.Matter Phys. \textbf{6}, 411 (2015) and references therein.

\bibitem{SebastianRepProgPhys2012} S. E. Sebastian, N. Harrison, and G. G.
Lonzarich, Rep. Prog. Phys. \textbf{75}, 102501 (2012).

\bibitem{ProustComptesRendus2013} Baptiste Vignolle, David Vignolles,
Marc-Henri Julien, and Cyril Proust, Comptes Rendus Phys. \textbf{14}, 39
(2013).

\bibitem{SebastianPhilTrans2011} Suchitra E. Sebastian, Neil Harrison and
Gilbert G. Lonzarich, Phil. Trans. R. Soc. A \textbf{369}, 1687 (2011).

\bibitem{HelmNd2009} T. Helm, M. V. Kartsovnik, M. Bartkowiak, N. Bittner,
M. Lambacher, A. Erb, J. Wosnitza, and R. Gross, Phys. Rev. Lett. \textbf{103%
}, 157002 (2009).

\bibitem{HelmNd2010} T. Helm, M. V. Kartsovnik, I. Sheikin, M. Bartkowiak,
F.Wolff-Fabris, N. Bittner, W. Biberacher, M. Lambacher, A. Erb, J. Wosnitza
et al., Phys. Rev. Lett. \textbf{105}, 247002 (2010).

\bibitem{HelmNd2011} M.V. Kartsovnik,T. Helm, C. Putzke, F.Wolff-Fabris, I.
Sheikin, S. Lepault, C. Proust, D. Vignolles, N. Bittner,W. Biberacher et
al., New J. Phys. \textbf{13}, 015001 (2011).

\bibitem{HelmNd2015} T. Helm, M. V. Kartsovnik, C. Proust, B. Vignolle, C.
Putzke, E. Kampert, I. Sheikin, E.-S. Choi, J. S. Brooks, N. Bittner, W.
Biberacher, A. Erb, J. Wosnitza, and R. Gross, Phys. Rev. B \textbf{92},
094501 (2015).

\bibitem{BaFeAs2011} Taichi Terashima, Nobuyuki Kurita, Megumi Tomita,
Kunihiro Kihou, Chul-Ho Lee, Yasuhide Tomioka, Toshimitsu Ito, Akira Iyo,
Hiroshi Eisaki, Tian Liang, Masamichi Nakajima, Shigeyuki Ishida, Shin-ichi
Uchida, Hisatomo Harima, and Shinya Uji, Phys. Rev. Lett. \textbf{107},
176402 (2011).

\bibitem{Graf2012} D. Graf, R. Stillwell, T. P. Murphy, J.-H. Park, E. C.
Palm, P. Schlottmann, R. D. McDonald, J. G. Analytis, I. R. Fisher, and S.
W. Tozer, Phys. Rev. B. \textbf{85}, 134503 (2012).

\bibitem{ColdeaReview2013} Amalia I. Coldea, Daniel Braithwaite, Antony
Carrington, Comptes Rendus Physique \textbf{14}, 94 (2013).

\bibitem{FeSeTerashima2014} T. Terashima, N. Kikugawa, A. Kiswandhi, E.-S.
Choi, J. S. Brooks, S. Kasahara, T. Watashige, H. Ikeda, T. Shibauchi, Y.
Matsuda, T. Wolf, A. E. B\"{o}hmer, F. Hardy, C. Meingast, H. v. L\"{o}%
hneysen, M.-T. Suzuki, R. Arita, and S. Uji, Phys. Rev. B. \textbf{90},
144517 (2014).

\bibitem{FeSeAudouard2015} A. Audouard, F. Duc, L. Drigo, P. Toulemonde, S.
Karlsson, P. Strobel, and A. Sulpice, Europhys. Lett. \textbf{109}, 27003
(2015).

\bibitem{FeSeWatsonPRB2015} M. D. Watson, T. K. Kim, A. A. Haghighirad, N.
R. Davies, A. McCollam, A. Narayanan, S. F. Blake, Y. L. Chen, S.
Ghannadzadeh, A. J. Schofield, M. Hoesch, C. Meingast, T. Wolf, and A. I.
Coldea, Phys. Rev. B \textbf{91}, 155106 (2015).

\bibitem{FeSeMQOPRL2015} M.\thinspace D. Watson, T. Yamashita, S. Kasahara,
W. Knafo, M. Nardone, J. Beard, F. Hardy, A. McCollam, A. Narayanan,
S.\thinspace F. Blake, T. Wolf, A.\thinspace A. Haghighirad, C. Meingast,
A.\thinspace J. Schofield, H. v. L\"{o}hneysen, Y. Matsuda, A.\thinspace I.
Coldea, and T. Shibauchi, Phys. Rev. Lett. \textbf{115}, 027006 (2015).

\bibitem{SebastianNature2008} Suchitra E. Sebastian, N. Harrison, E. Palm,
T. P. Murphy, C. H. Mielke, Ruixing Liang, D. A. Bonn, W. N. Hardy and G. G.
Lonzarich, Nature 454, 200 (2008)

\bibitem{AudouardPRL2009} Alain Audouard, Cyril Jaudet, David Vignolles,
Ruixing Liang, D. A. Bonn, W. N. Hardy, Louis Taillefer, and Cyril Proust,
Phys. Rev. Lett. \textbf{103}, 157003 (2009).

\bibitem{SingletonPRL2010} John Singleton, Clarina de la Cruz, R. D.
McDonald, Shiliang Li, Moaz Altarawneh, Paul Goddard, Isabel Franke, Dwight
Rickel, C. H. Mielke, Xin Yao, and Pengcheng Dai, Phys. Rev. Lett. \textbf{%
104}, 086403 (2010).

\bibitem{SebastianPNAS2010} S. E. Sebastian, N. Harrison, M. M. Altarawneh,
C. H. Mielke, R. Liang, D. A. Bonn, W. N. Hardy, G. G. Lonzarich, Proc.
Natl. Acad. Sci. U.S.A. \textbf{107}, 6175 (2010).

\bibitem{SebastianPRB2010} Suchitra E. Sebastian, N. Harrison, P. A.
Goddard, M. M. Altarawneh, C. H. Mielke, Ruixing Liang, D. A. Bonn, W. N.
Hardy, O. K. Andersen, and G. G. Lonzarich, Phys. Rev. B \textbf{81}, 214524
(2010).

\bibitem{SebastianPRL2012} Suchitra E. Sebastian, N. Harrison, Ruixing
Liang, D. A. Bonn, W. N. Hardy, C. H. Mielke, and G. G. Lonzarich, Phys.
Rev. Lett. \textbf{108}, 196403 (2012).

\bibitem{SebastianNature2014} S.E. Sebastian, N. Harrison, F.F. Balakirev,
M.M. Altarawneh, P.A. Goddard et al., Nature \textbf{511}, 61 (2014).

\bibitem{ProustNatureComm2015} N. Doiron-Leyraud, S. Badoux, S. Rene de
Cotret, S. Lepault, D. LeBoeuf, F. Laliberte, E. Hassinger, B.J. Ramshaw,
D.A. Bonn, W.N. Hardy,R. Liang, J.-H.. Park, D. Vignolles, B. Vignolle, L.
Taillefer and C. Proust, Nature Comm. \textbf{6}, 6034 (2015).

\bibitem{FournierARPES2010} D. Fournier, G. Levy, Y. Pennec, J. L.
McChesney, A. Bostwick, E. Rotenberg, R. Liang, W. N. Hardy, D. A. Bonn, I.
S. Elfimov \& A. Damascelli, Nature Physics \textbf{6}, 905 (2010).

\bibitem{Borisenko2006} S. V. Borisenko, A. A. Kordyuk, V. Zabolotnyy, J.
Geck, D. Inosov, A. Koitzsch, J. Fink, M. Knupfer, B. Buchner, V. Hinkov, C.
T. Lin, B. Keimer, T. Wolf, S. G. Chiuzbaian, L. Patthey, and R. Follath,
Phys. Rev. Lett. \textbf{96}, 117004 (2006).

\bibitem{NPhysPereg} T. Pereg-Barnea, H.Weber, G. Refael and M. Franz,
Quantum oscillations from Fermi arcs. Nat. Phys. \textbf{6}, 44-49 (2010).

\bibitem{XRayScience2012} G. Ghiringhelli, M. Le Tacon, M. Minola, S.
Blanco-Canosa, C. Mazzoli, N. Brookes, G. De Luca, A. Frano, D. Hawthorn, F.
He et al., Science \textbf{337}, 821 (2012).

\bibitem{XRayNatPhys2012} J. Chang, E. Blackburn, A. Holmes, N. Christensen,
J. Larsen, J. Mesot, R. Liang, D. Bonn, W. Hardy, A. Watenphul et al., Nat.
Phys. \textbf{8}, 871 (2012).

\bibitem{XRayPRL2012} A. J. Achkar, R. Sutarto, X. Mao, F. He, A. Frano, S.
Blanco-Canosa, M. Le Tacon, G. Ghiringhelli, L. Braicovich, M. Minola, M.
Moretti Sala, C. Mazzoli, R. Liang, D. A. Bonn, W. N. Hardy, B. Keimer, G.
A. Sawatzky, and D. G. Hawthorn, Phys. Rev. Lett. \textbf{109}, 167001
(2012).

\bibitem{Xray2016} S. Gerber, H. Jang, H. Nojiri, S. Matsuzawa, H. Yasumura,
D. A. Bonn, R. Liang, W. N. Hardy, Z. Islam, A. Mehta, S. Song, M. Sikorski,
D. Stefanescu, Y. Feng, S. A. Kivelson, T. P. Devereaux, Z.-X. Shen, C.-C.
Kao, W.-S. Lee, D. Zhu, J.-S. Lee, Science \textbf{350}, 949 (2015); H.
Jang, W.-S. Lee, H. Nojiri, S. Matsuzawa, H. Yasumura, L. Nie, A. V.
Maharaj, S. Gerber, Y. Liu, A. Mehta, D. A. Bonn, R. Liang, W. N. Hardy, C.
A. Burns, Z. Islam, S. Song, J. Hastings, T. P. Devereaux, Z.-X. Shen, S. A.
Kivelson, C.-C. Kao, D. Zhu, J.-S. Lee, PNAS \textbf{113}, 14645 (2016).

\bibitem{NMR2011Wu} T. Wu, H. Mayaffre, S. Kr{\" a}mer, M. Horvatic, C.
Berthier, W. Hardy, R. Liang, D. Bonn, and M.-H. Julien, Nature (London) 
\textbf{477}, 191 (2011).

\bibitem{NMR2015Wu} T. Wu, H. Mayaffre, S. Kr{\" a}amer, M. Horvatic, C.
Berthier, W. Hardy, R. Liang, D. Bonn, and M.-H. Julien, Nat. Commun. 
\textbf{6}, 6438 (2015).

\bibitem{CDWSoundVelocity} David LeBoeuf, S. Kr{\" a}mer, W. N. Hardy,
Ruixing Liang, D. A. Bonn \& Cyril Proust, Nature Physics \textbf{9}, 79
(2013).

\bibitem{ElPocketHall2007} David LeBoeuf, Nicolas Doiron-Leyraud, Julien
Levallois, R. Daou, J.-B. Bonnemaison, N. E. Hussey, L. Balicas, B. J.
Ramshaw, Ruixing Liang, D. A. Bonn, W. N. Hardy, S. Adachi, Cyril Proust \&
Louis Taillefer, Nature \textbf{450}, 533 (2007).

\bibitem{BadouxHall2016} S. Badoux, W. Tabis, F. Laliberte, G.
Grissonnanche, B. Vignolle, D. Vignolles, J. Beard, D. A. Bonn, W. N. Hardy,
R. Liang, N. Doiron-Leyraud, Louis Taillefer \& Cyril Proust, Nature \textbf{%
531}, 210 (2016).

\bibitem{ElPocketSeebeck2010} J. Chang, R. Daou, Cyril Proust, David
LeBoeuf, Nicolas Doiron-Leyraud, Francis Laliberte, B. Pingault, B. J.
Ramshaw, Ruixing Liang, D. A. Bonn, W. N. Hardy, H. Takagi, A. B. Antunes,
I. Sheikin, K. Behnia, and Louis Taillefer, Phys. Rev. Lett. \textbf{104},
057005 (2010).

\bibitem{EfimovPRB2008} I. S. Elfimov, G. A. Sawatzky, and A. Damascelli,
Phys. Rev. B \textbf{77}, 060504(R) (2008).

\bibitem{GarciaNJP2010} David Garcia-Aldea and Sudip Chakravarty, New
Journal of Physics \textbf{12}, 105005 (2010).

\bibitem{HarrisonNJP2012} N. Harrison and S. E. Sebastian, New Journal of
Physics \textbf{14}, 095023 (2012).

\bibitem{HarrisonSciRep2015} N. Harrison, B. J. Ramshaw and A. Shekhter,
Scientific Reports \textbf{5}, 10914 (2015).

\bibitem{Briffa2015} A. K. R. Briffa, E. Blackburn, S. M. Hayden, E. A.
Yelland, M. W. Long, and E. M. Forgan, Phys. Rev. B \textbf{93}, 094502
(2016).

\bibitem{Yelland2008} E. A. Yelland, J. Singleton, C. H. Mielke, N.
Harrison, F. F. Balakirev, B. Dabrowski, and J. R. Cooper, Phys. Rev. Lett. 
\textbf{100}, 047003 (2008).

\bibitem{Bangura2008} A. F. Bangura, J. D. Fletcher, A. Carrington, J.
Levallois, M. Nardone, B. Vignolle, P. J. Heard, N. Doiron-Leyraud, D.
LeBoeuf, L. Taillefer, S. Adachi, C. Proust, and N. E. Hussey, Phys. Rev.
Lett. \textbf{100}, 047004 (2008).

\bibitem{TanPNAS2015} B. S. Tan, N. Harrison, Z. Zhu, F. Balakirev, B. J.
Ramshaw, A. Srivastava, S. A. Sabok-Sayr, B. Dabrowski, G. G. Lonzarich, and
Suchitra E. Sebastian, Proc. Natl. Acad. Sci. U.S.A. \textbf{112}, 9568
(2015).


\bibitem{CommentDoping1} Doping changes the size of unreconstructed
Fermi-surface considerably, giving a very large relative change of the size
of any small reconstructed Fermi-surface pockets.

\bibitem{DopingDependence2015} B. J. Ramshaw, S. E. Sebastian, R. D.
McDonald, James Day, B. S. Tan, Z. Zhu, J. B. Betts, Ruixing Liang, D. A.
Bonn, W. N. Hardy, N. Harrison, Science \textbf{348}, 317 (2015). See in
particular Fig. 2.

\bibitem{SO} M.V. Kartsovnik, P.D. Grigoriev, W. Biberacher, N.D. Kushch, P.
Wyder, Phys. Rev. Lett. \textbf{89}, 126802 (2002).

\bibitem{Shub} P.D. Grigoriev, Phys. Rev. B \textbf{67}, 144401 (2003).

\bibitem{RET} P.D. Grigoriev, A. A. Sinchenko, P. Lejay, A. Hadj-Azzem, J.
Balay, O. Leynaud, V. N. Zverev and P. Monceau, Eur. Phys. J. B \textbf{89},
151 (2016). In this case slow oscillations in bilayer rare-earth tellurides
at a single frequency were interpreted as coming from an interlayer coupling
but with no warping.

\bibitem{Dingle} R.B. Dingle, Proc. Roy. Soc. \textbf{A211,} 517 (1952).

\bibitem{Champel2001} T. Champel and V. P. Mineev, Phil. Magazine B \textbf{%
81}, 55 (2001).

\bibitem{WIPRB2012} P. D. Grigoriev, M. V. Kartsovnik, W. Biberacher, Phys.
Rev. B \textbf{86}, 165125 (2012).

\bibitem{CommentDFtz} Eq. (\ref{Angtz}) assumes that $t_{z}\left(\boldsymbol{%
k}_{\parallel }\right) \approx const$. When $t_{z}\left( \boldsymbol{k}%
_{\parallel }\right) $ has a strong angular dependence, Eq. (\ref{Angtz}) is
modified,\cite{Mark92,Bergemann,GrigAMRO2010} but remains an oscillating
function of $k_{F}c^{\star }\tan \theta $.

\bibitem{Yam} K. Yamaji, J. Phys. Soc. Jpn. \textbf{58}, 1520 (1989).

\bibitem{JETPL2017MQO} P.D. Grigoriev and T. Ziman, JETP Lett. \textbf{106}%
(6), in press (2017) [Pis’ma v ZhETF \textbf{106}(6), 349 (2017)].

\bibitem{Andersen1995} O. K. Andersen, A.I. Liechtenstein, O. Jepsen, F.
Paulsen, J. Phys. Chem. Solids 56, 1573 (1995).

\bibitem{GrigMogInPlaneMR} P.D. Grigoriev and T.I. Mogilyuk, to be published.

\bibitem{Commentm} We take $m_{\beta }^{\ast }\approx 3.8m_{e}$ as obtained
from dHvA measurements for $\beta $-frequency,\cite{SebastianNature2008}
because within our model $m_{\alpha }^{\ast }\approx 1.6m_{e}$ is not a true
effective electron mass but rather an effective parameter coming from the
temperature dependence of the square of Dingle temperature (see Sec.IIC).

\bibitem{CommentFitRT} An excellent fit of experimental data by the
temperature damping factor $R_{T}$ of the Lifshitz-Kosevich formula has been
achieved (see Fig. 6c in Ref. \cite{AnnuReviewYBCO2015}), at least for one
doping level $p\approx 0.108$ over a wide temperature interval $1K<T<16$K.
Such a good fit is atypical even for materials which are much cleaner and
have stoichiometric chemical composition and almost ideal crystal structure.
In any case, the temperature-dependent Dingle factor due to e-e interactions
in YBCO should spoil the ideal dependence $R_{T}(T)$, which in Fig. 6c in
Ref. \cite{AnnuReviewYBCO2015} survived even up to $T=16$K. Therefore, we
believe that the quality of the fit may be accidental.

\bibitem{CommentTD} The electron level broadening $\Gamma $ is related to
the Dingle temperature $T_{D}=\Gamma /k_{B}$ and to the effective electron
mean free time $\tau =\hbar /2\Gamma $. Usually, at low temperatures one
takes into account only electron scattering by impurities, which gives a
temperature-independent Dingle factor. The contribution to $T_{D}$ from
electron-electron (e-e) and electron-phonon (e-ph) increases with
temperature and may be confused with the damping factor coming from
temperature $R_{T}$, thus leading to the incorrect estimate of the effective
mass from the oscillations. While in the lowest order of perturbation theory
the electron-phonon contributions to $T_{D}$ and to $m^{\ast }$ were shown%
\cite{Engelsberg1970,Shoenberg} to compensate each other, the
electron-electron contribution to $T_{D}$ in disordered metals needs further
theoretical study.

\bibitem{Engelsberg1970} S. Engelsberg, G. Simpson, Phys. Rev. B \textbf{2},
1657 (1970).

\bibitem{Mirlin} Y. Adamov, I. V. Gornyi, and A. D. Mirlin, Phys. Rev. B 
\textbf{73}, 045426 (2006).

\bibitem{CommentChemPotOsc} For simplicity we assume the chemical potential $%
\mu $ to be constant. In the opposite limit of constant particle density the
chemical potential is an oscillating function of magnetic field,\cite%
{Pavel2,ChampelChemPotOsc} which gives additional nonlinearity to the
functional $M[\rho (\varepsilon ,B)]$.

\bibitem{Mark92} M. V. Kartsovnik, V. N. Laukhin, S. I. Pesotskii, I. F.
Schegolev, V. M. Yakovenko, J. Phys. I \textbf{2}, 89 (1992).

\bibitem{Bergemann} C. Bergemann, S. R. Julian, A. P. Mackenzie, S.
NishiZaki, and Y. Maeno, Phys. Rev. Lett. \textbf{84}, 2662 (2000).

\bibitem{GrigAMRO2010} P.D. Grigoriev, Phys. Rev. B \textbf{81}, 205122
(2010).

\bibitem{Pavel2} P. Grigoriev, JETP \textbf{92}, 1090 (2001) [Zh. Eksp.
Teor. Fiz. \textbf{119}(6), 1257 (2001)].

\bibitem{ChampelChemPotOsc} Thierry Champel, Phys. Rev. B \textbf{64},
054407 (2001).
\end{thebibliography}
\end{document}